\RequirePackage{lineno}
\documentclass[final,3p,times]{elsarticle}
\usepackage{rotating}
\usepackage{epsfig} 
\usepackage{hyperref}
\usepackage{color}
\usepackage{lineno}
\usepackage{url}
\usepackage{multirow}

\usepackage{amsmath}
\usepackage{graphicx}
\usepackage[caption=false]{subfig}

\graphicspath{ {./plot/}, {./image/} }
\DeclareGraphicsExtensions{ .pdf, .png}

\makeatletter
\renewcommand{\p@subsection}{}
\renewcommand{\p@subsubsection}{}
\makeatother

\makeatletter
\let\LN@equation\equation
\let\LN@endequation\endequation
\renewcommand{\equation}{\linenomath\LN@equation}
\renewcommand{\endequation}{\LN@endequation\endlinenomath}
\let\LN@gather\gather
\let\LN@endgather\endgather
\renewcommand{\gather}{\linenomath\LN@gather}
\renewcommand{\endgather}{\LN@endgather\endlinenomath}
\makeatother

\newcommand{\snn}{\sqrt{s_{\textsc{nn}}}}
\newcommand{\qcd}{{\textsc{qcd}}}
\newcommand{\avfd}{{\textsc{avfd}}}
\newcommand{\cme}{{\textsc{cme}}}
\newcommand{\poi}{{\textsc{poi}}}
\newcommand{\rp}{{\textsc{rp}}}
\newcommand{\ep}{{\textsc{ep}}}
\newcommand{\Ru}{$^{96}_{44}$Ru}
\newcommand{\Zr}{$^{96}_{40}$Zr}
\newcommand{\au}{{\text{Au}}}
\newcommand{\auau}{{\text{AuAu}}}
\newcommand{\ruru}{{\text{RuRu}}}
\newcommand{\zrzr}{{\text{ZrZr}}}
\newcommand{\iso}{{\text{isobar}}}
\newcommand{\bkg}{{\text{Bkg}}}
\newcommand{\dg}{\Delta\gamma}
\newcommand{\fcme}{f_{\cme}}
\newcommand{\Ps}{\epsilon}
\newcommand {\mean}[1]   {\langle{#1}\rangle}

\begin{document}


\title{Revisit the Chiral Magnetic Effect Expectation in Isobaric Collisions at the Relativistic Heavy Ion Collider}

\author{Yicheng Feng}
\address{Department of Physics and Astronomy, Purdue University, West Lafayette, Indiana 47907, USA}

\author{Yufu Lin}
\address{Key Laboratory of Quark and Lepton Physics (MOE) and Institute of Particle Physics, Central China Normal University, Wuhan, Hubei 430079, China}

\author{Jie Zhao}
\address{Department of Physics and Astronomy, Purdue University, West Lafayette, Indiana 47907, USA}

\author{Fuqiang Wang\corref{cora}}
\cortext[cora]{Corresponding author}
\ead{fqwang@purdue.edu}
\address{Department of Physics and Astronomy, Purdue University, West Lafayette, Indiana 47907, USA}
\address{School of Science, Huzhou University, Huzhou, Zhejiang 313000, China}

\date{\today} 


\begin{abstract}
Isobaric \Ru+\Ru\ and \Zr+\Zr\ collisions at $\snn=200$~GeV have been conducted at the Relativistic Heavy Ion Collider to circumvent the large flow-induced background in searching for the chiral magnetic effect (\cme), predicted by the topological feature of quantum chromodynamics (QCD). 
Considering that the background in isobar collisions is approximately twice that in Au+Au collisions (due to the smaller multiplicity) and the \cme\ signal is approximately half (due to the weaker magnetic field), we caution that the \cme\ may not be detectable with the collected isobar data statistics, within $\sim$2$\sigma$ significance, if the axial charge per entropy density ($n_5/s$) and the QCD vacuum transition probability are system independent. This expectation is generally verified by the Anomalous-Viscous Fluid Dynamics (\avfd) model. While our estimate provides an approximate ``experimental'' baseline, theoretical uncertainties on the \cme\ remain large.
\end{abstract}


\maketitle

The chiral magnetic effect (\cme) in quantum chromodynamics (\qcd) refers to charge separation along strong magnetic field caused by imbalanced numbers of left-handed and right-handed quarks due to interactions of chiral fermions with metastable topological domains~\cite{Kharzeev:1998kz,Kharzeev:2007jp}. Such domains can form due to vacuum fluctuations in \qcd, which might provide a mechanism for the large CP violation in the early universe, believed to be responsible for the matter-antimatter asymmetry today~\cite{RevModPhys.76.1}. 
The \cme\ is being actively pursued in relativistic heavy ion collisions~\cite{Kharzeev:2015znc,Zhao:2018skm,Zhao:2019hta,Kharzeev:2019zgg,Li:2020dwr}. A common variable used to measure the \cme-induced charge separation in those collisions is the azimuthal correlator~\cite{Voloshin:2004vk}, $\dg=\gamma_{\textsc{os}}-\gamma_{\textsc{ss}}$ with $\gamma\equiv\mean{\mean{\cos(\phi_\alpha+\phi_\beta)}}$, where $\phi$ is the particle azimuthal angle relative to the reaction plane (\rp), and the subscript $\alpha$ and $\beta$ indicate two particles with either opposite charge sign (\textsc{os}) or same charge sign (\textsc{ss}). The measurements of $\dg$ in relativistic heavy ion collisions are contaminated by a major source of background, caused by correlations of particles from a common source (such as resonance decay) which itself is anisotropically distributed about the \rp\ as quantified by the elliptic flow parameter $v_2$~\cite{Wang:2009kd,Bzdak:2009fc,Schlichting:2010qia}. To circumvent the large flow background, many innovative approaches have been proposed~\cite{Adamczyk:2013kcb,Schukraft:2012ah,Zhao:2017nfq,Xu:2017qfs,Tang:2019pbl}. In addition, isobaric \Ru+\Ru\ and \Zr+\Zr\ collisions were proposed~\cite{Voloshin:2010ut} where the backgrounds are expected to be the same because of the same mass number of these isobars and the magnetic fields generated by these collisions (and hence the \cme\ signals) are appreciably different because of the different atomic numbers.

Isobar collisions at nucleon-nucleon center-of-mass energy of $\snn=200$~GeV have been conducted at the Relativistic Heavy Ion Collider (RHIC)~\cite{Skokov:2016yrj} and approximately $2\times10^9$ good minimum bias  collision events~\cite{STAR:2019bjg} were collected by the STAR experiment for each of the isobar collision systems. The CME signal is expected to be significant only in mid-central collisions and vanishes in peripheral and central collisions where the magnetic field disappears. A back-of-the-envelope estimation of the significance of \cme\ measurement is as follows. 

\hangindent=\parindent 
\underline{Statistical uncertainty.}
A single pair quantity $\cos(\phi_{\alpha}+\phi_{\beta})$ is an approximately uniform distribution between $[-1,1]$ with a root-mean-square of $\sigma_1=2/\sqrt{12}$. The typical charged hadron multiplicity measured in the STAR Time Projection Chamber ($|\eta|\lesssim1$) in 20-60\% centrality Au+Au collisions is of the order of $N_{\poi}^{\auau}\approx320$~\cite{Abelev:2008ab}, all of which are used in analysis as particles of interest (\poi). Since multiplicity scales with $A$, the mass number of the colliding nuclei, $N_{\poi}^{\iso}\approx160$ in isobar collisions of the same centrality range. Thus the width of the event-wise $\mean{\cos(\phi_{\alpha}+\phi_{\beta})}$ distribution is $\sigma_N=\sqrt{2}\sigma_1/\sqrt{(N_{\poi}^{\iso}/2)^2}\approx0.01$. Experimentally, the \rp\ is assessed by the reconstructed event plane (\ep) and corrected by the \ep\ resolution factor $R_{\ep}\approx0.5$~\footnote{In experimental analysis, the \ep\ is often reconstructed using the so-called subevent method, and the \poi's used for $\dg$ are reduced in statistics. We use all particles in our estimation so that the estimated uncertainties are the best achievable. The \ep\ in such analysis is often assessed by a single particle with the \ep\ resolution being the particle $v_2$. In this case the overall resolution effect on the final statistical uncertainty is given by $\sqrt{N}v_2$ under small $v_2$ approximation~\cite{Poskanzer:1998yz}. This factor is typically also of the order of 0.5.}. The statistical uncertainty on the $\dg$ difference between the isobar collision systems is therefore $\sigma=\sqrt{2}\sigma_N/\sqrt{N_{\rm evt}^{\iso}}/R_{\ep}\approx10^{-6}$ (where $N_{\rm evt}^{\iso}\approx 10^9$ is the number of events of each species in the 20-60\% centrality range).

\hangindent=\parindent
\underline{Signal strength.}
To estimate the \cme\ signal strength in isobar collisions, we use Au+Au collisions as a guide. The inclusive $\dg$ measured in 20-60\% central Au+Au collisions is on the order of $\dg^{\auau}\approx2\times10^{-4}$~\cite{Abelev:2009ac,Abelev:2009ad}. Since it is dominated by the flow-induced background and thus approximately scales inversely with $N$ (because the number of background correlation sources is likely proportional to $N$ and $\dg$ is a pair-wise quantity, while the $v_2$ parameters are expected to be similar between isobar and Au+Au collisions), $\dg^{\iso}\approx4\times10^{-4}$ is expected in isobar collisions. Suppose the \cme\ signal fraction in isobar collisions is $\fcme\equiv\dg_{\cme}/\dg\sim10\%$, similar to what the Au+Au data~\cite{Zhao:2018blc,Zhao:2020utk,STAR:2021pwb} seem to indicate, then the \cme\ signal difference between the isobar systems is on the order of $\Delta(\dg_{\cme}^{\iso})\approx 0.15\fcme\dg^{\iso}\approx 6\times10^{-6}$ where the factor 0.15 comes from the magnetic field difference~\footnote{Although the atomic numbers differ by 10\% between the isobars \Ru\ and \Zr, the expected $\Delta B/B$ difference is a bit smaller because of a slight radii difference in the charge density distributions~\cite{Deng:2016knn,Shi:2017cpu,Xu:2017zcn}. Since $\dg_{\cme}\propto B^2$~\cite{Kharzeev:2007jp,Kharzeev:2015znc}, the effect is $2\Delta B/B$. This uncertainty on $\Delta B/B$ does not significantly alter our estimates.}.

\noindent
The above back-of-the-envelope estimation suggests that the signal $\Delta(\dg_{\cme}^{\iso})$ is 6 times the expected statistical uncertainty $\sigma$, i.e.~a measurement of $6\sigma$ significance (systematic uncertainty is expected to be small~\cite{STAR:2019bjg}). This is consistent, within a factor of 2, with the more sophisticated estimate by Deng et al.~\cite{Deng:2016knn} of $5\sigma$ with an assumed $\fcme=1/3$ and a factor of 5 smaller data volume, or $3.4\sigma$ using the numbers in our estimation.

In those estimates the crucial input is the \cme\ signal fraction $\fcme$ for isobar collisions. 
{\em It has been generally expected that the \cme\ fractions are similar in Au+Au and isobar collisions.}
However, there is no reason to expect so. 
In fact, as aforementioned, the background scales with $1/A$, which yields a factor of 2 larger background in isobar collisions than in Au+Au.  The magnetic field in heavy ion collisions is expected to scale as $B\propto A^{1/3}$ and since $\dg_{\cme}\propto B^2$, it would possibly lead to a factor of 1.6 smaller signal in isobar collisions.
In other words, based on simple reasoning, {\em the natural expectation is rather $\fcme\propto A^{5/3}$.}
As a net result, there could possibly be over a factor of 3 difference in the $\fcme$ between isobar and Au+Au collisions, smaller for the former.
This suggests that our back-of-the-envelope estimate would give $2\sigma$ significance and the more sophisticated estimate~\cite{Deng:2016knn} would give $1\sigma$ significance.
Clearly, a more rigorous investigation is needed, for which we use the Anomalous-Viscous Fluid Dynamics (\avfd)~\cite{Jiang:2016wve,Shi:2017cpu}.

The \avfd\ model implements anomalous fluid dynamics to describe the evolution of fermion currents in the quark-gluon plasma created in relativistic heavy ion collisions~\cite{Jiang:2016wve,Shi:2017cpu,Shi:2019wzi}. The underlying bulk medium evolution is described by the VISH2+1 hydrodynamics~\cite{Heinz:2015arc}. The model integrates the anomalous fluid dynamics with the normal viscous hydrodynamics  in the same framework. It incorporates ingredients including the initial conditions, the magnetic fields, and the viscous transport coefficients, and allows interplay between the evolution of the axial charge current and the bulk medium.
In our simulation we used the version EBE-\avfd\ Beta1.0 which includes event-wise fluctuations in the initial conditions~\cite{Shi:2017cpu,Shi:2019wzi}.

The \cme\ arises from the finite axial charge current due to the imbalanced numbers of left-handed and right-handed quarks. The magnitude of the axial charge per entropy density ($n_5/s$) is, however, rather poorly known~\cite{Kharzeev:2004ey,Muller:2010jd,Shi:2017cpu}. 
The $n_5/s$ value is taken as an input to \avfd.
We can, however, use the available Au+Au data as a benchmark to calibrate \avfd. 
To that end we simulate  isobar  as well as Au+Au collisions by \avfd. 
While the \cme\ signals  in individual collision systems are difficult to gauge theoretically (and also experimentally at present), their relative strengths in isobar collisions vs.~Au+Au collisions should be more robust.

\begin{figure*}
    \centering
    \includegraphics[width=0.325\textwidth]{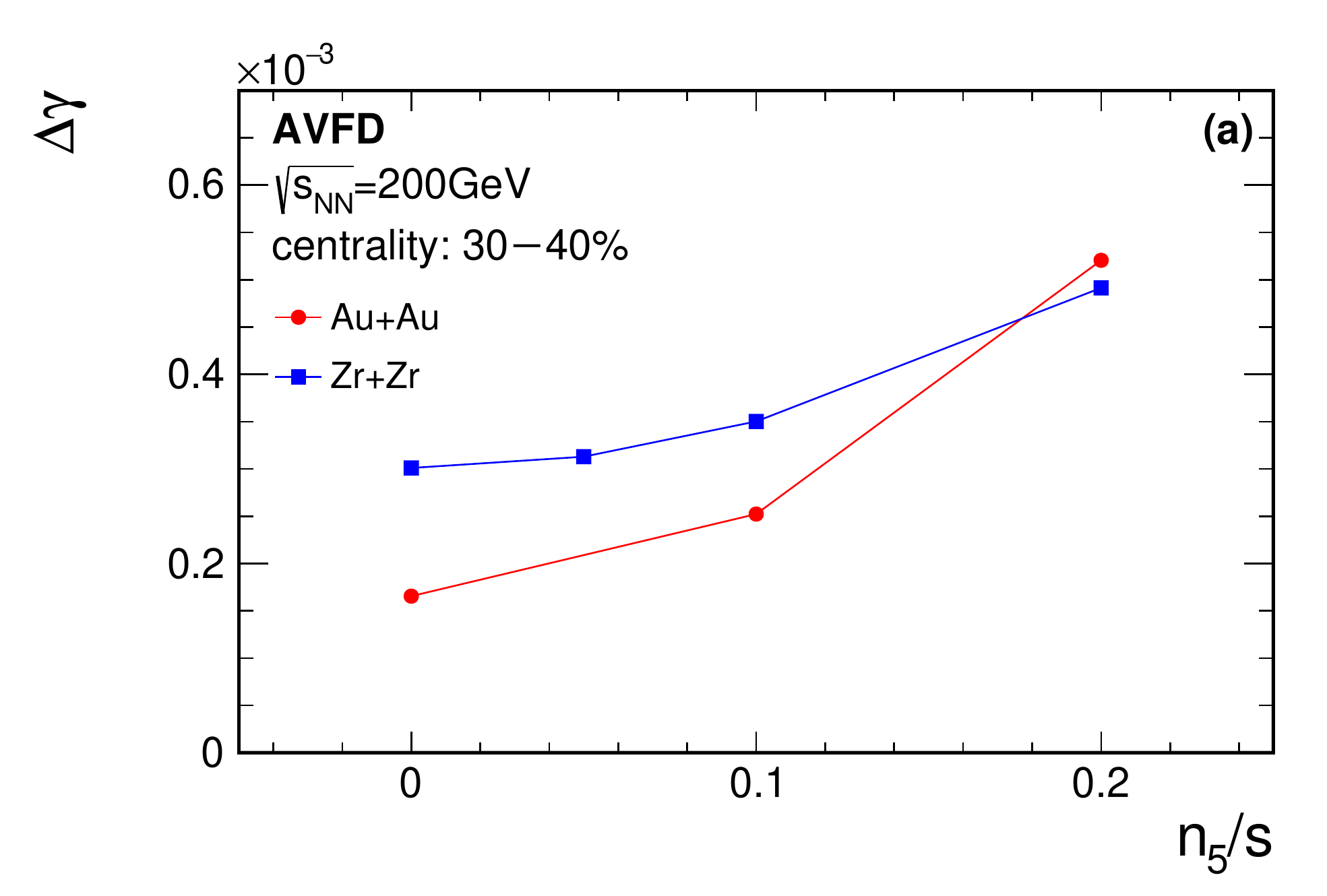}
    \includegraphics[width=0.325\textwidth]{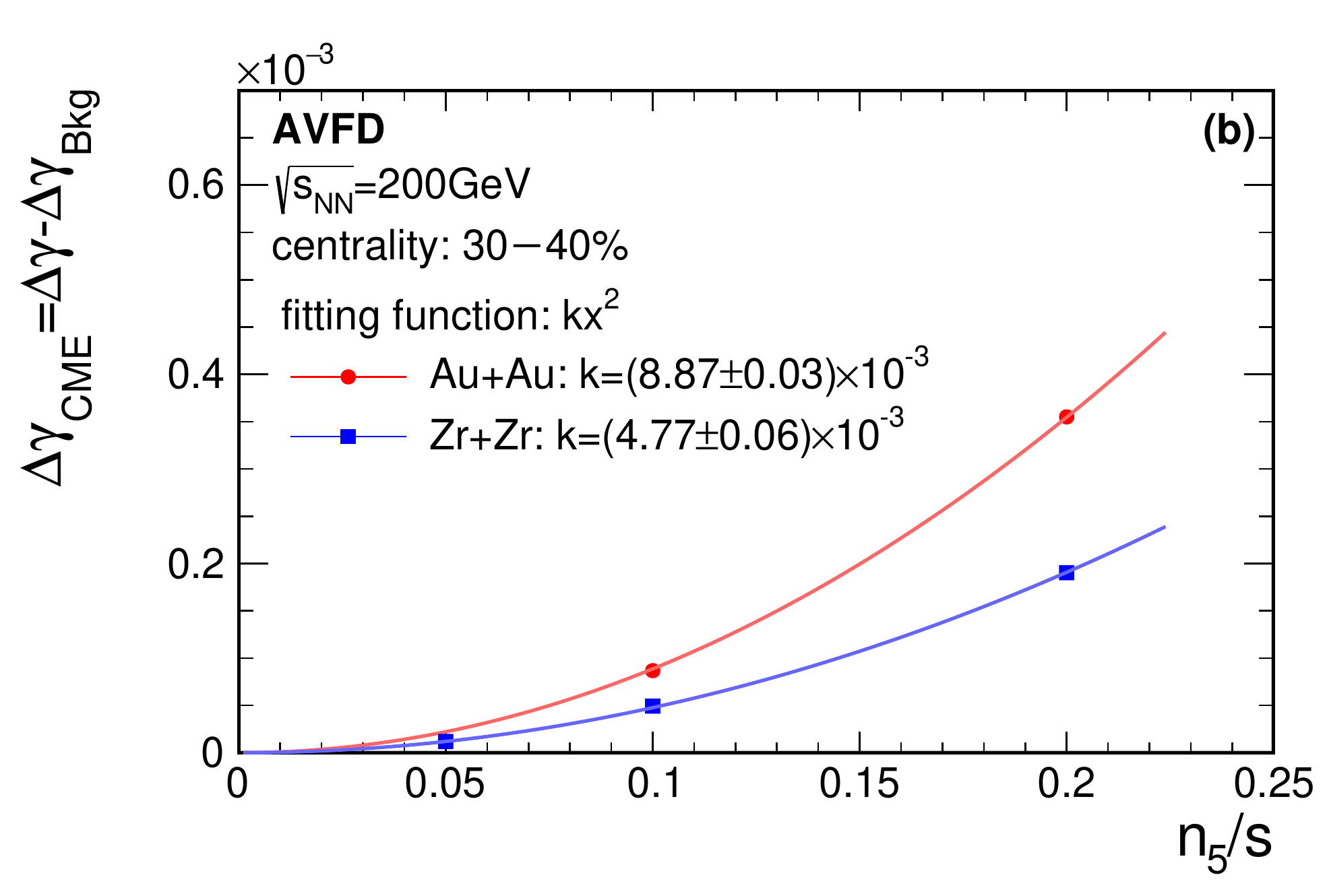}
    \includegraphics[width=0.325\textwidth]{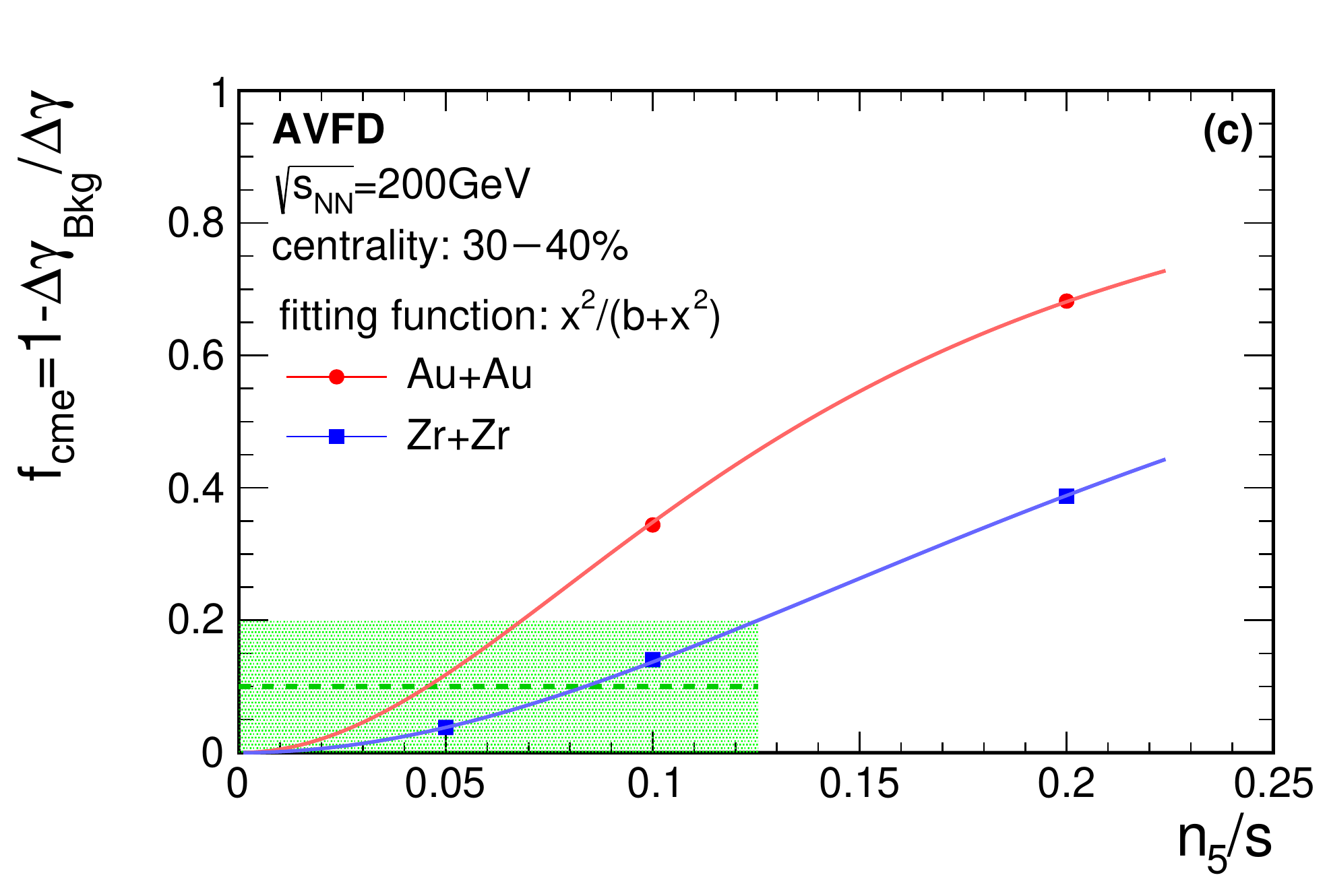}
    \caption{(a) inclusive $\dg$ calculated by \avfd\ in 30-40\% Au+Au and Zr+Zr collisions at $\snn=200$~GeV as functions of the axial charge per entropy density $n_5/s$; (b) background-subtracted \cme\ signal $\dg_{\cme}$, where the background is taken as the $\dg$ at $n_5/s=0$; (c) the \cme\ signal fraction $\fcme$, where the curves are fits to $\frac{(n_5/s)^2}{(n_5/s)^2+b}$ with $b$ related to the background-to-signal ratio. The horizontal line at $\fcme=10\%$ and the shaded area are to guide the eye. Statistical uncertainties are too small to be visible.}
    \label{fig}
\end{figure*}

Figure~\ref{fig}(a) shows the $\dg$ calculated by \avfd\ in 30-40\% centrality Au+Au and Zr+Zr collisions at $\snn=200$~GeV. The Ru+Ru results are similar to Zr+Zr. 
The leftmost data points at $n_5/s=0$ are entirely due to flow background.
The backgrounds differ by approximately a factor of 1.9 between Zr+Zr and Au+Au; this is consistent with the aforementioned multiplicity dilution.
We can extract the \cme\ signals $\dg_{\cme}$ at $n_5/s\ne0$ by subtracting the flow background $\dg_{\bkg}$ taken as the $\dg$ at $n_5/s=0$. The $\dg_{\cme}$ is shown in Fig.~\ref{fig}(b) as function of $n_5/s$.
It is quadratic, $\dg_{\cme}=k(n_5/s)^2$, as expected because $\dg$ is a two-particle correlation variable.
The signal strengths differ by also approximately a factor of 1.9 at the {\em same} $n_5/s$ value between Au+Au and Zr+Zr collisions, but in the opposite direction of the background difference. This is not unexpected because 
the initial magnetic field strengths in \avfd\ differs by a factor of 1.7 between Au+Au and Zr+Zr collisions~\cite{Shi:2017cpu}, somewhat larger than the aforementioned $A^{1/3}$ scaling.
This would result in a factor of $2.9$ difference in $\dg_{\cme}$. 
However, there could be final-state effect reducing the \cme\ signal~\cite{Ma:2011uma} and this reduction would be stronger in Au+Au than isobar collisions.
Also, it is possible that the \cme\ signal could be somewhat diluted by multiplicity (similar to the background) if it arises from multiple independent domains of axial charges in local magnetic fields. In other words, the \cme\ signal could behave more like nonflow, rather than flow as normally expressed by the $a_1$ parameter in a Fourier series $1+2a_1\sin(\phi)+2v_2\cos(2\phi)+...$, where $\dg_{\cme}=2a_1^2$ and $a_1\propto n_5/s$~\cite{Shi:2017cpu}. This would result in a larger multiplicity dilution in Au+Au than isobar collisions.
Thus, it may not be entirely unexpected that the relative \cme\ signal in Au+Au with respect to isobar collisions is smaller than the $B^2$ scaling; \avfd\ indicates a relative reduction factor of $\Ps^{\iso}/\Ps^{\auau}\approx2.9/1.9$ (here $0<\Ps^{\auau}, \Ps^{\iso}<1$). Moreover, the initial temperature is expected to be higher in Au+Au than isobar collisions, which would lead to a larger sphaleron transition probability and hence a larger initial \cme\ signal. So the final-state reduction factor could be even larger. 
Nevertheless, it is evident that the \avfd\ results generally support the estimates by the aforedescribed  simple reasoning. 

The \cme\ fraction, $\fcme$, is shown in Fig.~\ref{fig}(c) for both Zr+Zr and Au+Au collisions. It is now not a surprise, as noted above, that the \cme\ fractions are not the same between isobar and Au+Au collisions at a given $n_5/s$.
Using shorthand notation $r_{\bkg}\equiv\dg_{\bkg}^{\iso}/\dg_{\bkg}^{\auau}$ ($\approx 1.9$) for the background ratio and $r_{\cme}\equiv k^{\iso}/k^{\auau}$ ($\approx 1/1.9$) for the ratio of the proportionality coefficient $k$, it is straightforward to show
\begin{equation}
\fcme^{\iso}=\frac{\fcme^{\auau}}{\fcme^{\auau}+r(1-\fcme^{\auau})}\,,
\end{equation}
where 
$r=\frac{r_{\bkg}}{r_{\cme}}\cdot\frac{(n_5/s)^2_{\auau}}{(n_5/s)^2_{\iso}}$.
At small $\fcme$, the \cme\ fraction in isobar collisions is a factor of $r$ smaller than that in Au+Au collisions. At the same $n_5/s$, this factor is $r\approx 3.6$ in \avfd; this implies that the isobar data may only yield a 1.7$\sigma$ measurement (recall that it would be $6\sigma$ if $r=1$). 

Although one would naively expect the $n_5/s$ to be similar between isobar collisions and Au+Au collisions, as we have assumed above, it is not inconceivable that they could differ.
In fact, by using the initial collision geometries and the glasma picture with a fixed gluon satuaration scale ($Q_s^2=1.25$~GeV$^2$), the authors of \avfd\ estimate that the $n_5/s$ in isobar collisions is about 1.5 times that in Au+Au collisions, both for the 30--40\% centrality~\cite{Shi:2017cpu}. This would yield $r_{\cme}\approx1.5^2/1.9\approx1.2$, i.e.~slighly larger \cme\ signal strength $\dg_{\cme}$ in isobar than Au+Au collisions, or only a factor of $1.2/1.9\approx0.6$ smaller in $\fcme$. It would, in turn, yield a \cme\ signal difference measurement of 3.8$\sigma$ significance between the isobar collision systems. However, the expected higher initial temperature in Au+Au than isobar collisions (i.e.~a larger sphaleron transition probability) would reduce this significance value. 

Taking $r_{\bkg}\approx\frac{A^{\au}}{A^{\iso}}$ and $r_{\cme}\approx\left(\frac{A^{\iso}}{A^{\au}}\right)^{2/3}\cdot\frac{\Ps^{\iso}}{\Ps^{\auau}}$, 
a pocket formula for the significance of \cme\ measurement in isobar collisions may be written as
\begin{equation}
    \frac{\Delta(\dg_{\cme}^{\iso})}{\sigma}\approx
    \frac{\sqrt{3N_{\rm evt}^{\iso}}}{2}R_{\textsc{EP}}^{\iso}
    \left(N_{\poi}^{\auau}\fcme^{\auau}\dg^{\auau}\right)
    \left(\frac{\Ps^{\iso}}{\Ps^{\auau}}\right)
    \left(\frac{A_{\iso}}{A_{\au}}\right)^{5/3}
    \left[\frac{(n_5/s)_{\iso}}{(n_5/s)_{\auau}}\right]^2
    \left(\frac{S^{\iso}}{S^{\auau}}\right)
    \left(\frac{\Delta B}{B}\right)_{\iso}\,,
    \label{eq}
\end{equation}
where $N_{\rm evt}^{\iso}$ is the number of isobar collision events of each species in the considered centrality range, 
$R_{\textsc{EP}}^{\iso}$ is the \ep\ resolution, and $(\Delta B/B)_{\iso}$ is the relative magnetic field difference between Ru+Ru and Zr+Zr collisions.
In Eq.~(\ref{eq}), we have also included for completeness the effect of different sphaleron transition probabilities on the $\dg_{\cme}$ observable, $S^{\iso}/S^{\auau}$, which is largely uncertain theoretically.

We emphasize that our estimation exploits the relative comparisons between isobar and Au+Au collisions. The multiplicity scaling of the background should be relatively robust at a given energy. We note, however, that the multiplicity scaling cannot hold over a wide range of collision energies because the background contributions, such as resonances and jets, do not necessarily scale with multiplicity at different energies. The relative strength of the magnetic fields should also be robust, although their absolute magnitudes and lifetimes are largely uncertain which are encoded in the resultant \cme\ signal strengths. The uncertain elements of our estimation are primarily four. 
(i) The $\fcme^{\auau}$ measurements presently have relatively large uncertainties~\cite{Zhao:2018blc,Zhao:2020utk,Adam:2020zsu,STAR:2021pwb,Feng:2021pgf,Sirunyan:2017quh,Acharya:2017fau}; all measurements seem to suggest that it is smaller than 20\%. We used $\fcme^{\auau}\sim10\%$ in our estimate.
(ii) The relative axial charge amplitude is probably difficult to pin down theoretically. We used $\frac{(n_5/s)_{\iso}}{(n_5/s)_{\auau}}\sim$1 (as a guess) and 1.5 (as from \avfd~\cite{Shi:2017cpu}).
(iii) The relative sphaleron transition probability of the QCD vacuum is probably also difficult to determine theoretically; we did not explicitly consider this. 
(iv) The \cme\ reduction factor in the final state is also uncertain; \avfd\ indicates $\Ps^{\iso}/\Ps^{\auau}\approx2.9/1.9$ which has been implicitly included in our estimate. Only the magnetic field difference is considered in obtaining this reduction factor; if the sphaleron transition probability is included in \avfd, then this reduction factor value would be larger. 

We have so far focused on the signal difference between isobar Ru+Ru and Zr+Zr collisions. The backgrounds in these two systems are similar but not identical. Energy density functional theory calculations of the \Ru\ and \Zr\ nuclei indicate that the eccentricities of isobar collisions can differ by relatively 2-3\%~\cite{Xu:2017zcn,Li:2018oec}. If the $\fcme$ in isobar collisions is similar to that in Au+Au, say $\fcme\sim 10\%$~\cite{Zhao:2018blc,Zhao:2020utk}, then the signal-to-background ratio in the isobar \cme\ signal difference $\Delta(\dg^{\iso})$ would be $\sim0.15\fcme/0.03=1/2$, still much better than the signal-to-background ratio of 1/10 in $\dg^{\auau}$ in Au+Au collisions. If, however, the isobar \cme\ signal is reduced by a factor of $r\approx3.6$ compared to Au+Au, then the signal-to-background ratio for $\Delta(\dg^{\iso})$ would be $\sim$1/7, not much better than that for $\dg^{\auau}$ in Au+Au collisions.

The background difference between the isobar Ru+Ru and Zr+Zr collisions can be well gauged by the relative elliptic flow measurement ($r_{v_2}$). To cancel the background entirely, one would take the difference $\dg^{\ruru}-r_{v_2}\dg^{\zrzr}$. Suppose $r_{v_2}\approx1.03$, then the relative difference would be $0.12\fcme$ instead of $0.15\fcme$. The effect is small, reducing the significance by 20\% or so.

In summary, the \cme\ fraction $\fcme$ in the azimuthal correlator $\dg$ measurement could be a factor of $\sim$3--4 smaller in isobar collisions compared to that in Au+Au collisions by considering the relative background contributions and magnetic field strengths. If $\fcme\sim 10\%$ in Au+Au collisions, then the present isobar collision statistics may give an 1--2$\sigma$ effect, not enough to identify a possible \cme\ signal. If the sphaleron transition probability is smaller in isobar collisions because of the likely lower initial temperature achieved in those collisions than in Au+Au, then the significance would be even weaker.
If, on the other hand, the $n_5/s$ is a factor of 1.5 larger in isobar collisions than in Au+Au, as the glasma estimate in \avfd\ suggests, then a 3--4$\sigma$ significance could be expected. These are ballpark estimates with inevitably sizeable uncertainties, and being from easy-to-understand, back-of-the-envelope estimations, can serve as good benchmarks. The ultimate answer, obviously, has to come from the experimental isobar collision data themselves. 

We thank Dr.~James Dunlop, Dr.~Evan Finch, Dr.~Jinfeng Liao, Dr. Hanlin Li, Dr.~Shuzhe Shi, Dr.~Aihong Tang, Dr.~Gang Wang, and Dr.~Haojie Xu for fruitful discussions. We thank Dr.~Shuzhe Shi and Dr.~Jinfeng Liao for providing the EBE-\avfd\ Beta1.0 source code. 
This work is supported in part by the U.S.~Department of Energy (Grant No.~DE-SC0012910), the National Natural Science Foundation of China (Grant Nos.~12035006, 12075085, 12047568), the Fundamental Research Funds for the Central Universities (Grant No.~CCNU19ZN019), the Ministry of Science and Technology of China (Grant Nos.~2016YFE0104800, 2020YFE0202001), and the China Scholarship Council.

\bibliographystyle{unsrt} 
\bibliography{../rc/ref}

\end{document}